\documentclass[structabstract]{aa}
\usepackage{graphicx}
\usepackage{soul}
\usepackage{txfonts}
\usepackage[usenames,dvipsnames]{color}
\usepackage{supertabular}
\usepackage{longtable}
\usepackage{natbib}
\bibpunct{(}{)}{;}{a}{}{,}

\begin{document}
\def\mycolumnwidth{8.0cm}
\def\myhalfcolumnwidth{3.84cm}
\def\myseventhcolumnwidth{1.04cm}
\def\myttcolumnwidth{5.20cm}
\def\mytncolumnwidth{0.32cm}
\titlerunning{Hot Jupiters with TTVs in $Kepler$ data}
\title{Multiple planets or exomoons in $Kepler$ hot Jupiter systems with transit timing variations?}

\authorrunning{R. Szab\'o et al.}
\author{
R. Szab\'o$^{1}$,
Gy. M. Szab\'o$^{1,2,3}$,
G. D\'alya$^{4}$,
A. E. Simon$^{1,2}$, 
G. Hodos\'an$^{1,4}$,
L. L. Kiss$^{1,2,5}$
}
\institute{
MTA CSFK, Konkoly-Thege Mikl\'os \'ut 15-17, H-1121 Budapest, Hungary \email{rszabo@konkoly.hu}
\and 
ELTE Gothard--Lend\"ulet Exoplanet Research Group, H-9704 Szombathely, Szent Imre herceg \'ut 112, Hungary
\and
Dept. of Exp. Physics \& Astronomical Observatory,
      University of Szeged, H-6720 Szeged, Hungary
\and 
E\"otv\"os Lor\'and University,  P\'azm\'any P\'eter s\'et\'any 1/A, H-1117 Budapest, Hungary
\and 
Sydney Institute for Astronomy, School of Physics, University of Sydney, NSW 2006, Australia
}

\abstract
{}
{Hot Jupiters are thought to belong to single-planet systems. Somewhat surprisingly, some hot Jupiters have been reported to exhibit transit timing variations (TTVs). The aim of this paper is to identify the origin of these observations, identify possible periodic biases leading to false TTV detections, and refine the sample to a few candidates with likely dynamical TTVs.} 
{We present TTV frequencies and amplitudes of hot Jupiters in $Kepler$ Q0--6 data with Fourier analysis and a frequency-dependent bootstrap calculation to assess the false alarm probability levels of the detections.} 
{We identified 36 systems with TTV above four standard deviation confidence, about half of them exhibiting multiple TTV frequencies.
Fifteen of these objects (\object{HAT-P-7b}, \object{KOI-13}, 127, 183, 188, 190, 196, 225, 254, 428, 607, 609, 684, 774, 1176) probably show TTVs due to a systematic observational effect: long cadence data sampling is regularly shifted transit-by-transit, interacting with the transit light curves, introducing a periodic bias, and leading to a stroboscopic period. For other systems, the activity and rotation of the host star can modulate light curves and explain the observed TTVs. By excluding the systems that were inadequately sampled, showed TTV periods related to the stellar rotation, or turned out to be false positives or suspects, we ended up with seven systems. Three of them (KOI-186, 897, 977) show the weakest stellar rotation features, and these are our best candidates for dynamically induced TTV variations.}
{Those systems with periodic TTVs that we cannot explain with systematics from observation, stellar rotation, activity, or inadequate sampling may be multiple systems or even exomoon hosts.}
\keywords{planetary systems -- stars: binaries: eclipsing -- techniques: photometric}

\date{Received / Accepted}

\maketitle

\section{Introduction}

Transit timing tariation (TTV) is a major diagnostics of various system parameters of extrasolar planets
\citep{holman05,agol05}. In multiplanet systems, planets perturb each other, leading to correlated TTVs of them \citep{holman10, lissauer11}. TTVs have also uncovered the presence of further non-transiting planets in planetary systems 
\citep{ballard11, ford12b}.

According to our current view, hot Jupiters occur as single planets, since they have been not detected in multiplanet systems. This picture suggests that hot Jupiters occupy unperturbed orbits, hence their orbital motion is Keplerian, and they exhibit strictly periodic transit times. In contrast to this picture, current literature reports a considerable number of hot Jupiters with TTV, which are often periodic \citep{steffen12a,ford12a}. 

In this work we publish TTV periods, TTV amplitudes, and significance levels for hot-Jupiter candidates in $Kepler$ data. 

Transit times covering Q0--Q6 were published in a catalog by \cite{ford12a}\footnote{http://www.astro.ufl.edu/$\sim$eford/data/kepler/} (see \cite{steffen12b} for more details),  whose data set has become the basis of several TTV studies). The main aims of this study are to critically revise the \cite{ford12a} catalog and to look for possible nondynamical processes that may cause virtual TTVs. To this end, we analyzed the stellar variations of the systems in the original $Kepler$ data up to Q9 besides exploring the catalog itself. The main conclusions of this study are the following.
\begin{enumerate}
\item{} Timing data of some already published systems with suspected TTV can be satisfactorily explained by nondynamical reasons, such as stroboscopic period due to even sampling or light curve distortions due to stellar activity;
\item{}  About 2\%{} of the Jupiter-size candidates passed all tests and show a TTV that presently has an unknown origin, and obviously needs further studies with follow-ups.
\item{} A fraction of systems with TTV signals tend to exhibit multiple TTV periods (confirmed by \cite{mazeh13}) that are incompatible with sampling or stellar rotation effects.
\end{enumerate}

We briefly introduce the most exciting systems and discuss the possible sources of TTVs for these planets.

\begin{figure*}
\centering\includegraphics[bb=75 450 740 640,width=12cm]{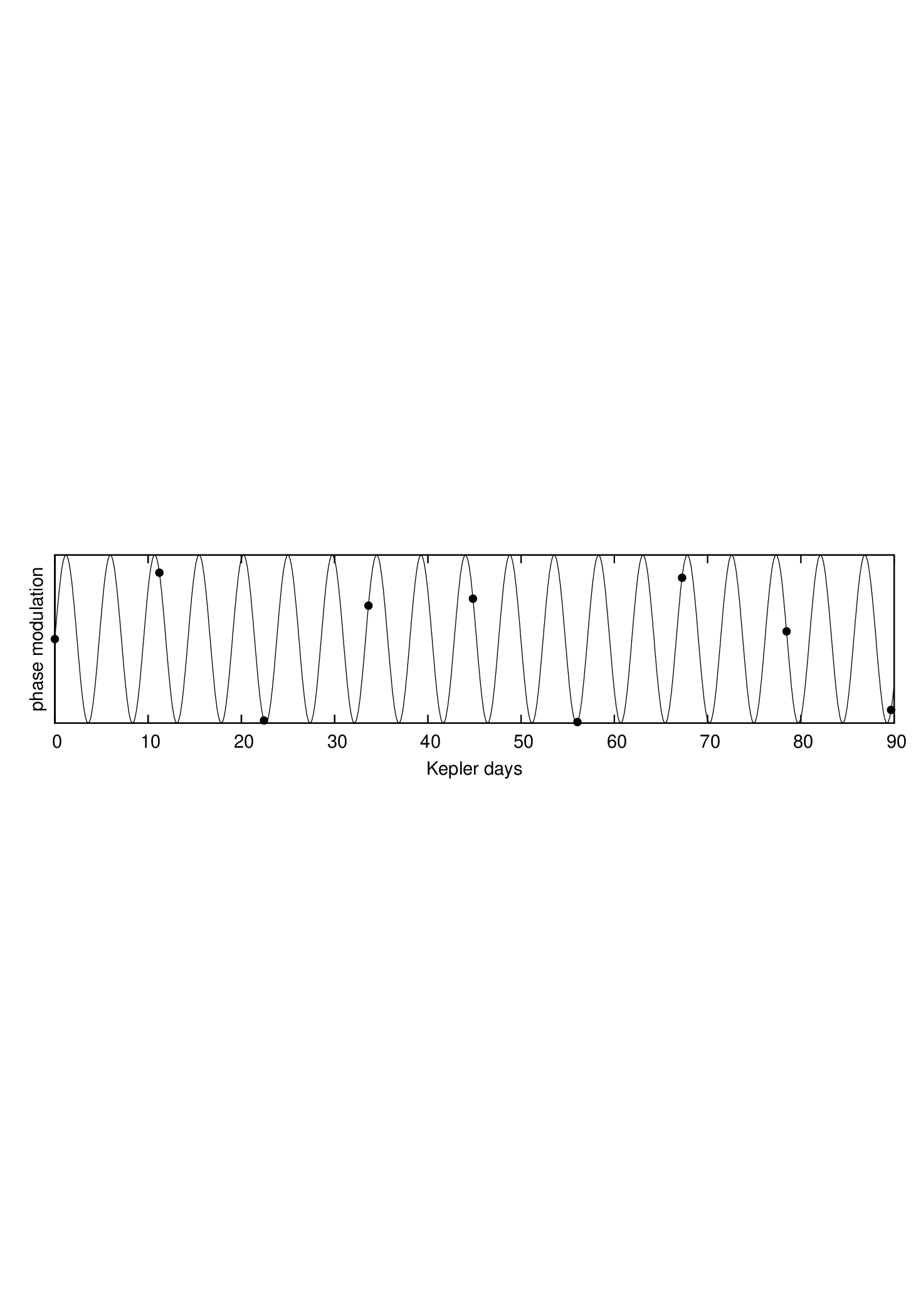}

\centering\includegraphics[bb=75 475 740 640,width=12cm]{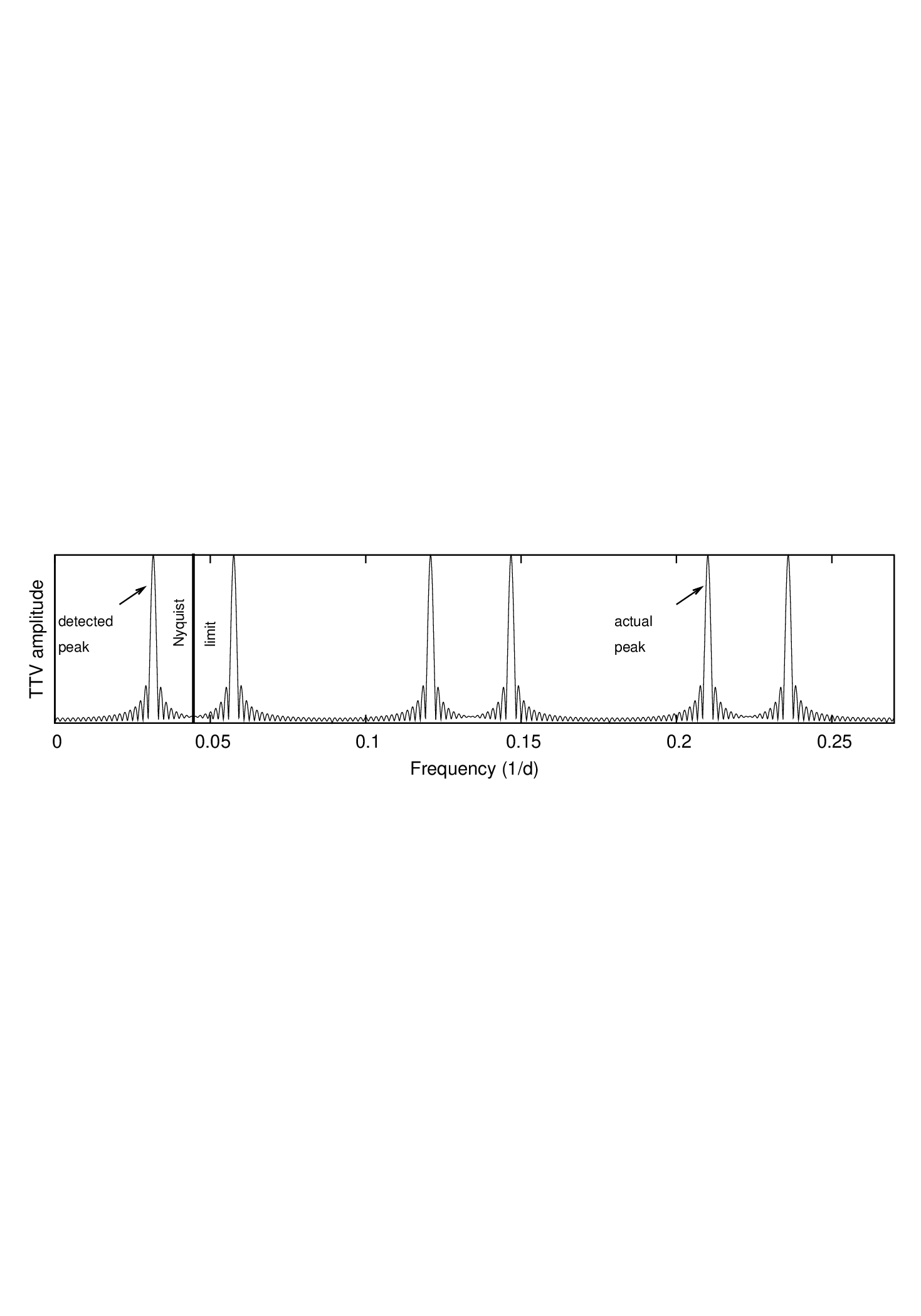}
\caption{Illustration of the detection of a super-Nyquist modulation in the orbital motion with Fourier analysis. Top panel: one quarter (90-d) long detail from a 600-d long simulation. Dots represent sampling by transits. Bottom panel: Fourier series of the 600 d long data set, extended beyond the Nyquist frequency. The actual frequency and the sub-Nyquist detections are highlighted. This specific model had an orbital (sampling) and modulation period of 11.21 and 4.76 days (which were taken from the ZIP code and altitude-above-sea-level of the Budapest station of Konkoly Observatory, as uncorrelated random numbers).}
\label{Nyquist}
\end{figure*}

\section{Data selection and analysis}

We selected all single companions from the second list of $Kepler$ planet candidates \citep{batalha12}, and filtered the list for planets larger than 6~$R_E$. Then all systems with incomplete data records, too few observed minima (we required more than 12 transits of a given system for our analysis) or inaccurate orbital periods were neglected. In total, 159 candidates were selected for further analysis. The median planet size and orbital periods were 11.23 $R_E$ and 5.70 days in the sample, and the longest orbital period was 54 days. After our vetting process the maximum orbital period became 8.36 days (sample of 19 candidates, see also Tables~\ref{table} and \ref{apptable}) and finally 3.2 days (sample of three systems), so that the reduced sample contains really close-in planets, mostly hot Jupiters. 

Transit times of these systems were taken from the TTV catalog of \cite{ford12a} and were processed by standard Fourier analysis 
with the {\tt MUFRAN} package \citep{kollath90}. Here we wish to illustrate the most serious drawback of applying the Fourier approach in such an analysis. Once the orbital phase is modulated by a periodic signal, a periodic TTV will be observed, that is, a periodic TTV confidently proves a periodic modulation of the orbital motion. However, the period of the modulation cannot be reconstructed from the TTV if it is shorter than two orbital periods. This is because transits undersample such frequent variations, and the observed frequency will be an alias of the super-Nyquist peak. This is illustrated in Fig.~\ref{Nyquist}. In the general case, this can lead to an order-of-magnitude difference between the detected and the actual period. This is unfavorable in the sense that one can practically tell nothing about the period of the TTV, and the detected period will only signify a lower 
frequency limit for a periodic process. However, the amplitude of the modulation can be properly reconstructed from TTV data (except strictly resonant cases), and the TTV amplitude can be involved in further analysis.

The Fourier transforms were evaluated individually. The presence of TTVs were characterized by the detected period, the TTV amplitude, and the height of the detected peak above the noise level. We considered the planet candidates with at least one peak higher than four standard deviations above the grass level as systems exhibiting periodic TTVs. We emphasize that we do not assume a normal distribution for the noise, since it is most probably non-Gaussian. Nevertheless, the standard deviation can be computed for any skewed distributions, but one has to be careful when interpreting the significance levels of the detections.

The reliability of the detected peaks was characterized with bootstrap FAP levels, calculated from the timing error in each transit as provided by the Ford et al. (2012) catalog. In the case of all systems, we simulated a large number (20,000) of artificial time series, containing no signal but the noise of the original measurement. (Artificial data points had the same time coordinates as original data, their value was normally distributed around zero mean, and the errors of the observed transit data points were the standard deviation of test data belonging to that time coordinate.) Test data were Fourier-transformed. Then, at each frequency, we observed the 0.9995 and 0.995 quantile of amplitudes, as indicators of the FAP levels belonging to the 0.05\%{} and 0.5\%{} levels. The distribution was slightly smoothed to remove the local numerical fluctuations but to conserve the general frequency dependence envelope of FAP levels. The highest peak exceeded the 0.5\%{} FAP level in the case of all candidates.

In the next step, we explored how observational and astrophysical artifacts can mimic TTV signals and whether such a bias acts in the case of our candidates. We identified two processes capable of enhancing the TTV signal without any dynamical reason, such as stroboscopic TTV due to sampling and stellar activity. We discuss them in the following.

\section{Nondynamical processes mimicking TTV signals}

\subsection{Stroboscopic frequencies due to regularly-spaced sampling}\label{strobo}

We found that in some cases, TTVs exhibited prominent frequencies that can be easily explained by sampling effects. Most of the data are long cadence, and they have a sampling rate of 29.424 minutes. This is in the order of the duration of one whole transit, resulting in three to six  points belonging to most of the transits. Since the orbital period is not an exact multiple of the cadence rate, the consecutive transits are sampled at different positions, but the lags will evolve very regularly, thanks to the even sampling rate. The instantly realized sampling can introduce biases in the transit time determination, and because of the regular evolution of the entire process, a virtual variation in transit times may be measured. (A similar feature is seen, e.g., in trailed CCD images of asteroids, where the very regular sampling results in a virtually undulating motion with a subpixel amplitude.)

\begin{figure}
\centering\includegraphics[width=8cm]{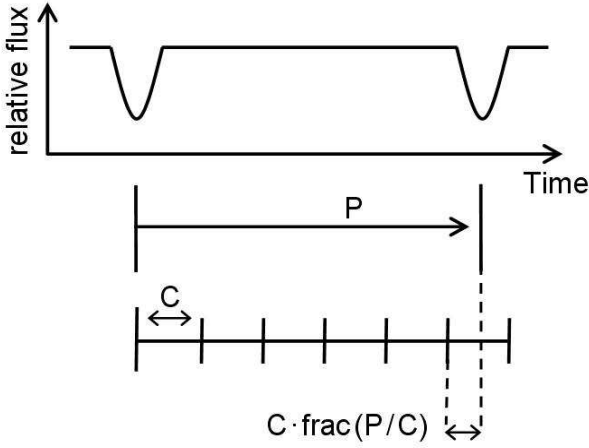}
\caption{Calculating the stroboscopic frequency in transit timings due to the uneven sampling of consecutive transits. 
$P$ and $C$ are the orbital period and the cadence, respectively, $frac$ means the fractional part.}
\label{stroboscopic}
\end{figure}

In Fig.~\ref{stroboscopic}. we illustrate the derivation of the stroboscopic frequency. Because the cadence of exposures is (obviously) not synchronized to transit
midtimes, a systematic lag will develop between the sampling structure of the consecutive transits. The sampling comb will be off one transit later by a time lag of $|P-nC|$, where $P$ and $C$ are the orbital period and the cadence rate, and $n$ is an appropriate integer. We know that $n=[P/C]_{-0}^{+1}$, where $[]$ means the integer part. There are two cases, one when $nC$ is slightly less than $P$ (by less then $P/2$, see this case in Fig.~\ref{stroboscopic}.), and another one, when $nC$ is slightly larger than $P$. In the two cases, denoting the fractional part with $][$ brackets, we can write 
\begin{equation}
P-\left[ {P\over C}\right] C = C\left( {P\over C} - \left[ {P\over C}\right] \right) = C \left] {P\over C}\right[,
\end{equation}
and following the same logic and inverting the sign because the quantity in absolute value brackets is negative
\begin{equation}
\left| P - \left[ {P\over C} +1 \right] C \right| = C \left] 1- {P\over C}\right[ .
\end{equation}
Thus, the time lag can be written as $C\cdot s$ if we define $s=\min(]P/C[, 1-]P/C[)$. The $C\cdot s$ time lag is $s$ times one cadence, and after $1/s$ occurrences (i.e transits), the initial cadence-to-transit configuration will be repeated. This means that the stroboscopic peak has $P/s$ period, hence $s/P$ frequency. Because $s$ is always less than $0.5$ by definition, the stroboscopic peak will always emerge, but will not always be high enough to be detected. In summary, if a transiting system exhibits a TTV period near $P/s$, this period should be considered to be suspicious, because it may refer to the stroboscopic period of the system. We eliminated fifteen objects (\object{HAT-P-7b}, \object{KOI-13}, 127, 183, 188, 190, 196, 225, 254, 428, 607, 609, 684, 774, 1176) owing to stroboscopic TTV periods in this step.  
Some frequencies in the spectra of the remaining candidates were also found to be stroboscopic. 

\begin{figure}
\centering\includegraphics[bb=190 190 478 540,height=8cm,angle=270,clip=T]{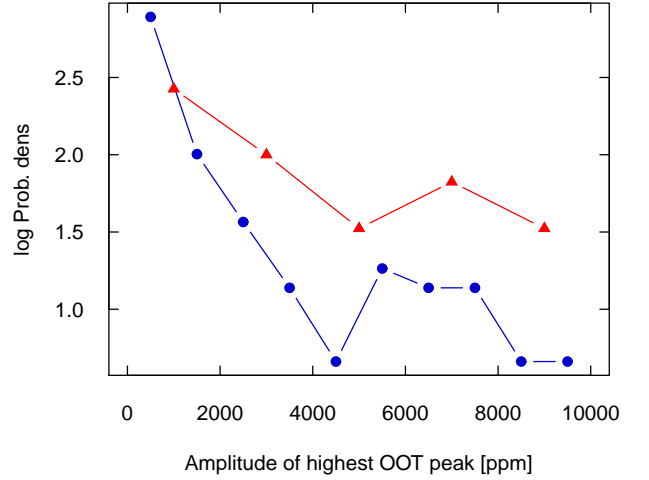}
\caption{Distribution of the amplitude of the largest peak in the out-of-transit periodogram of the 159 Jupiter-sized candidates (blue dots, lower curve) and the TTV suspect sample consisting of 36 members (red triangles, upper curve).}
\label{relampl}
\end{figure}

\begin{center}
\begin{table*}
\caption[]{Periodic TTV detections above the level of 4 standard deviations that survive our sequence of rigorous tests. Boldface denotes our three best candidates. Periods in italic are ambiguous based on the analysis of the optical light curves of the candidates. Stroboscopic periods are omitted.} \label{table}
\begin{tabular}{rlllllrrrcc}
\hline \hline
\noalign{\smallskip}
KOI &  $R/R_E$    & P (d)     &$M_*$& $T_{eff}$ & $R_*$ & TTV per. (d) &  Amp. (min) & Sign. lev. (s.t.d.) & Notes \\
\noalign{\smallskip}
\hline
\noalign{\smallskip}
      131.01  &  9.61  & 5.0142325 & 1.13 & 6244 & 1.21  & 114.377216  &0.6192 & 4.2  &  1\\
 {\bf 186.01}  & 12.35  & 3.2432603 & 1.06 & 5826 & 0.97  &  13.877132  &0.3888 & 6.1  &    \\ 
               &        &           &      &      &       &  7.2431227  &0.3744 & 5.8  & \\
 256.01  & 25.34  & 1.3786789  & 0.65 & 3639&  0.52 &  41.755397  &0.4464 & 4.4 &  2 \\
 823.01  &  7.89  & 1.028414   &1.1  & 5976  &0.96  &  {\it 142.897970} &{\it 2.2608} &{\it 18.8} & \\
                                          &&&&&&       {\it  59.477785} &{\it 1.6272} &{\it 13.0} & \\                                      
                                          &&&&&&        48.562549 &1.2960 &10.0  &  \\
                                          &&&&&&        84.061870 &1.1952 & 9.0  &  \\
                                          &&&&&&       343.28870  &1.0800 & 8.0  &  \\
                                          &&&&&&        19.033480 &0.8784 & 6.2  &  \\
 882.01  & 12.17  & 1.9568102  &0.93 & 5081  &$0.79^a$  &   41.879554 &0.5328 & 6.4  &  2 \\
 {\bf 897.01}  & 12.41  & 2.0523497  &1.07 & 5734  &1.03  &   81.973932 &0.3888 & 4.7  & 2  \\
 {\bf 977.01}  & 63.45  & 1.3537763  &0.21 &4204   &16.48 &  101.522843 &12.2832 & 5.6  & 3 \\
&& &&&&                                                       20.029644 & 9.6768 & 4.0 &\\
\noalign{\smallskip}
\hline
\end{tabular}
\tablefoot{(1): \cite{santerne12}, \cite{prsa11}, (2): \cite{batalha12} V-shaped (3): \cite{ford11} phase linked variations, 
a:  During the revision of this work slightly different stellar radii have become  available in the MAST database 
for most of our candidates, but we decided to stick to the values published earlier for two reasons: first, in order to have consistent masses that were not updated and second, because the changes are small and do not influence our conclusions in any way. The only exception is KOI-882, which had a radius of 0.55 ${\rm R_{\sun}}$ previously.}
\end{table*}
\end{center}

\begin{figure*}
\centering\includegraphics[height=18.5cm,angle=270]{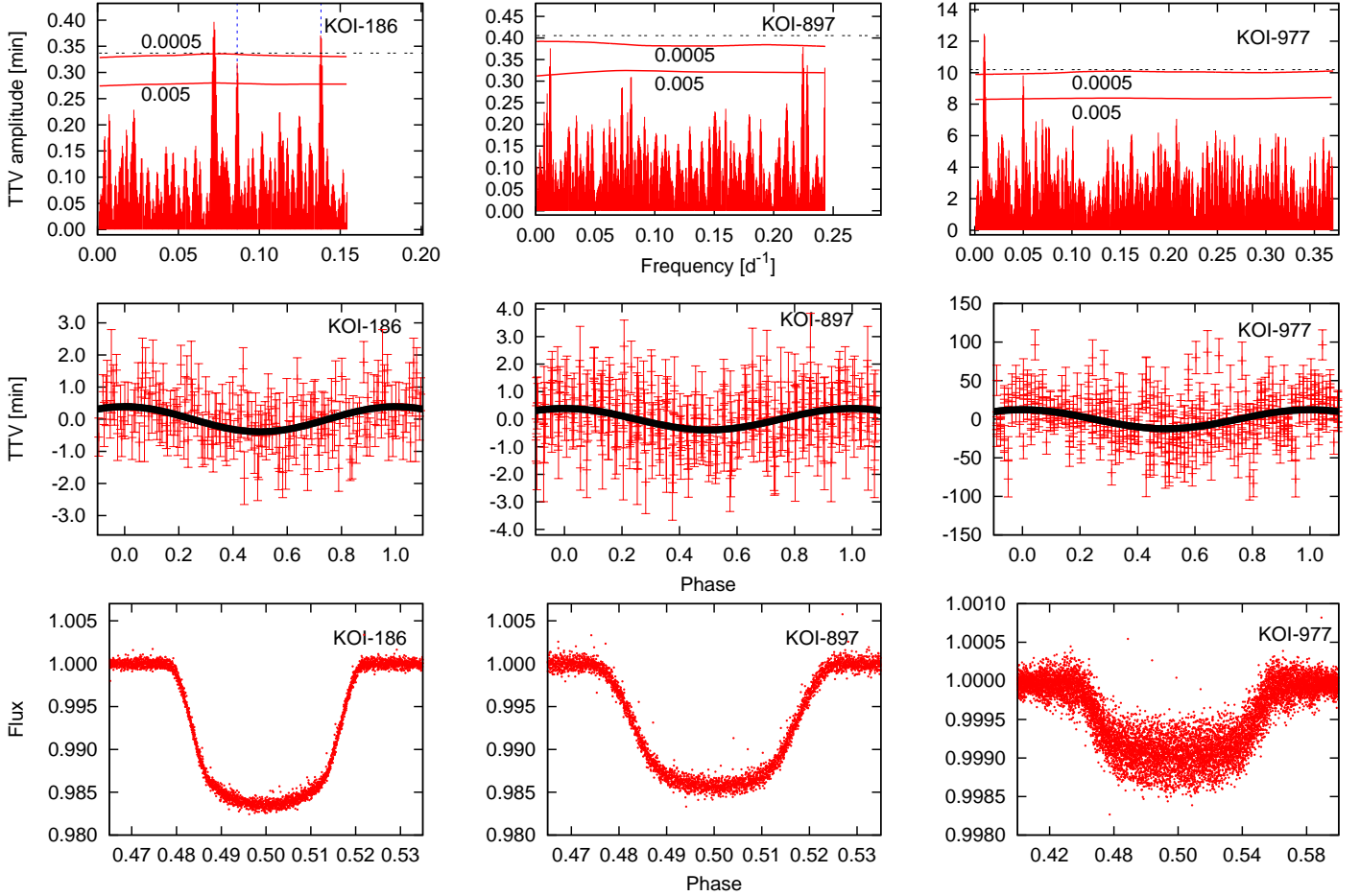}
\caption{{\bf Upper panels:} Fourier spectra of the transit timing variation of our best candidates. The horizontal dashed lines show the 5 s.t.d detection level, while the curves denote the frequency dependent, bootstrap false alarm probability (fap) of peak detections. Spurious peaks are denoted by vertical dashed lines in the case of KOI-186. {\bf Middle panels:} Folded TTV curves by the best periods. In case of KOI-977 the highest, 101.5 day peak was used. {\bf Lower panels:}  Transit shapes of the best candidates.}
\label{faps}
\end{figure*}

\subsection{Stellar activity}\label{activity}

Stellar activity is a known source of TTV, because star spots can modulate light curve shape, hence periodically modulate the transit times. Obviously, the effect highly depends on the actual algorithm of transit time fitting, thus in our special case, on the algorithm used to build the Ford et al. catalog. Finding and testing a transit time-fitting algorithm that is optimized to a stability against stellar activity is beyond the scope of our paper, and we are in no position to check the lowest activity level that can still bias the algorithms in the current $Kepler$ pipeline. We will assume, however, that any activity with uncommon level may be a source of an artificial TTV signal. Because this activity level will be a strong alert against a dynamically excited TTV, we suggest simply rejecting such hot-Jupiter candidates from the list of dynamically induced TTV candidates.

Almost all of our stars in the TTV sample show some variability due to rotation and/or spots in their light curves.
To quantify what fraction of the TTV candidate sample is contaminated, we explored the out-of-transit light variation, and registered the amplitude and the frequency of the highest peak in the Fourier transform. This is a proxy of stellar activity, and it also reports about any pulsation, systematics, etc., that are also able to affect transit times. The sample of TTV suspects was compared to the sample of all hot Jupiters. If the bias caused by the activity is negligible, the distribution of frequencies and amplitudes will be statistically indistinguishable in the two samples.

\subsection{Possible influence of stellar activity on TTVs}

To test whether there is a systematic bias in the TTV-positive sample toward more expressed stellar rotation signal because of higher activity, we compared the out-of-transit signals of the TTV-positive sample to the set of all single Jupiter-sized planet candidates. After stitching sections together in Q8--Q9 data where there are no prominent gaps, and internal shifts are minimal, transits and secondary eclipses were masked out, and a high-pass filter with a characteristic frequency at ${\rm 70\thinspace d^{-1}}$ was applied to remove long-period signals. Then we calculated a Fourier transform of the data series. The out-of-transit variations were characterized by the height of the highest frequency peak that emerged in the processed data set.

The amplitude of the highest frequency peak is compared in Fig.~\ref{relampl} in the case of TTV positive and all hot Jupiters. The figure is log-scaled, to emphasize the differences that mostly emerge in the wings of distributions. Evidently, the two samples are different for out-of-transit variations, and stellar rotation is expressed more in a subset of the TTV-positive list. The smallest amplitude peaks were observed at around 300-500 ppm amplitude in the large sample, while three of our TTV-positive candidates had peak amplitudes below 400 ppm (these are: KOI-186, 897, and 977), which are therefore the best TTV-positive candidates, those least affected by stellar rotation effects. 

We note that the frequency of the out-of-transit peak is widely separated from the TTV frequency in practically all cases of the TTV-positive hot Jupiters with expressed stellar rotation (see the discussions for each individual target in Sec.~\ref{individual}). Therefore, one cannot make a direct link between TTVs and stellar rotation. However, based on the amplitude distributions, an indirect connection seems to exist. In summary, we vetted three TTV candidates (KOI-412, 822, 895) solely on the fact that their transit fits may be influenced by the rotational light variation of the host star. Some other stars showing strong rotational signs were found to be false positives or eclipsing binary systems (e.g. KOI-1003 and KOI-1152).

\section{Results}

Thirty-six systems out of the 159 were identified showing periodic TTVs. Fifteen detections showed stroboscopic periods only (Sec.~\ref{strobo}), and two additional systems were excluded from the beginning, because the TTV period was found to agree with the rotation period of the host star: (KOI-883) and \object{Kepler-17b} (formerly KOI-203, \cite{desert11}). Of the remaining candidates, 19 candidates show TTV peaks that are convincingly far from the stroboscopic periods and also they were not suggested to be blended objects. We introduce these systems in the following section.

We found that the TTV periods of three more objects are related to the stellar rotation period (Sec.~\ref{activity}) based on their {\it Kepler} light curves (KOI-412, 822, 895). During the revision of this paper, some of these candidates were also rejected as false positives or reclassified as eclipsing binaries. These rejected objects from the sample of 19 candidates are listed in Table~\ref{apptable}. Until this writing, there are seven TTV-positive $Kepler$ targets that have still survived as planet candidates.
These are listed in Table~\ref{table}. The tables contain stellar and planetary parameters taken from the PlanetQuest website\footnote{http://planetquest.jpl.nasa.gov/kepler/}. These are compatible with the values from \cite{borucki11}, the Kepler Input Catalog \citep{brown11}, and \cite{batalha12}. In \cite{batalha12}, however, the KIC values were improved by matching them to the Yonsei-Yale stellar evolution models \citep{demarque04,steffen13}. 

Out of the seven candidates we found that three systems, namely KOI-186, 897, and 977, are the best ones based on their clear separation of the frequencies resulting from presumably stellar variations and other remaining TTV periodicities. The frequency spectrum of the TTVs of our best candidates are shown in the upper panels of Fig.~\ref{faps}. The detection level for five standard deviations (s.t.d.) and the false alarm probabilities for two levels (0.005 and 0.0005, the latter corresponding to our 5 s.t.d.) are also shown. We note that although it is more judicious to compute the false alarm probability levels to assess the significance of the individual peaks, the relative flatness of the fap levels indicates that the actual noise distributions are not far from the Gaussian in our cases. 

We also plotted the folded TTV curves with the fitted periodicities in the middle panels. In the case of KOI-977 we plotted the fit with the 101.5-day periodicity. To assess the reliability of the period detections of our best candidates, we performed a likelihood ratio test, according to \cite{lupton93}. This test makes use of the deviance (often denoted as D) defined as 
\begin{equation}
$$D=-2 \log L_{0} - 2\log L_{model}$$, 
\end{equation}
where the $\log L$ terms represent the log-likelihood of the fitted null hypothesis and the model, respectively. D is compared to a $\chi^2$-distribution with a degree of freedom that is the difference of the degree of freedom (dof) of
the individual fits. In our case, the null hypothesis was the zero signal (having one parameter, i.e. the mean value), and the fitted model was the monoperiodic signal with three additional parameters (the assumed period of the TTV, and its best-fit amplitude and phase). Therefore the calculated $D$ statistics were compared to a $\chi^2$-distribution of 3 dof. The resulting $p$ parameters (showing how likely random fluctuations can mimic at least as strong a signal as we detected) were $p=0.0027$, $0.0004$, and $0.0002$ for KOI-186, KOI-897, and KOI-977, respectively, which is independent evidence 
of this periodicity in the measured TTVs. We note that the origin of the higher $p$ in the case of KOI-186 might be the presence of the two additional frequencies in the TTV spectrum (one stroboscopic and the other related to the stellar rotation).

Finally, in the lower panels of Fig.~\ref{faps} the transit shapes are plotted for our best candidates based on available Q0--Q6 data. The same detrending method was applied as in \cite{szabo11}. The orbital periods were taken from \cite{batalha12}. The large scatter in the case of KOI-977 is primarily due to the large size of the host star (for discussion see in Sec.~\ref{best}) and to the correspondingly smaller observable transit depth. We have found no indication of a possible diluted eclipsing binary scenario in these cases.

\subsection{Remarks on individual systems}\label{individual}

Most of our candidates in Tables~\ref{table} and \ref{apptable}. are listed in \cite{ford11}. We only refer to that paper in the tables if it provides relevant information in the context of this work. Surprisingly, a few of these systems exhibit multiple frequency peaks.

Of the systems listed below, KOI-189, 897, and 977 show the least prominent rotation effects, therefore these are our best candidates. The periods of the best candidates are 14, 82, and 101.5 days. Since the detected TTV period is only an upper limit for the actual period of the modulations, it may be that the actual periods are shorter than the detected ones (even an order-of-magnitude difference can easily be suspected). The amplitude of TTVs is less than one minute in two cases (KOI-186 and 897) and 12 minutes for KOI-977.

\subsubsection{Best candidates}\label{best}
 
\noindent {\bf  \object{KOI-186}} This Jupiter-sized planet orbits a solar-type star (M=1.06\thinspace ${\rm M_{\odot}}$). We found three significant periodicities, namely 13.877, 7.243 and 11.583 days. The last is found to be stroboscopic, while the second one is very close to the presumed stellar rotation period (7.84\thinspace d), so we are left with a TTV candidate featuring a single periodic variation with 13.877 days period. The star exhibits a low-amplitude (338\thinspace ppm), long-period, out-of-transit variation with a 94-day period, which is likely independent of the TTV signal.

\noindent {\bf \object{KOI-897}} This planetary candidate is slightly larger than Jupiter and orbits a close-to solar mass host star with a period of 2.68 days. There is a single period in its TTV Fourier-spectrum with 82.0 days, while another peak of 4.45\thinspace d is found to be stroboscopic. The highest peak out-of-transit has a relatively low amplitude of 338\thinspace ppm and a period of 115\thinspace d, making KOI-897 a serious target for a dynamically caused TTV.

\noindent {\bf \object{KOI-977}} The host star of this object has a mass of $0.21\thinspace {\rm M_{\odot}}$ and a radius of $16.48\thinspace {\rm R_{\odot}}$ with an effective temperature of 4204 K according to the Kepler Input Catalog. The nature of this star was corroborated by \cite{muirhead12}, who clearly identified it as a giant based on near-infrared spectroscopic observations. The radius of the companion is correspondingly large, $63.45\thinspace {\rm R_{E}}$, the largest in our sample. The mass estimate of the host star is obviously too low, most probably a result of the procedure that was optimized to select main-sequence targets for the $Kepler$ Mission.  Since the stellar parameters for KOI 977 seem suspicious, the hot Jupiter classification of this candidate may be in question. Two significant frequencies were found in the TTV spectrum at 101.5 and 20.0 days.

\subsubsection{Other candidates}

\noindent {\bf \object{KOI-131}} \cite{santerne12} observed this object as part of their large spectroscopic follow-up program 
of $Kepler$ planet candidates. They found a relatively fast-rotating exoplanet host star with $v\thinspace \rm{sin}\thinspace i$ = $27\pm$1 $\rm{km}\thinspace\rm{s}^{-1}$ and placed an upper limit of 14.3\thinspace ${\rm M_{J}}$
on the mass of the companion. The authors note that both the planetary and blend scenarios are compatible with their data. In addition, the 
object is listed in the $Kepler$ eclipsing binary catalog \citep{prsa11} with twice the orbital period. This is presumably because of the slightly different depth of the odd and even minima based on the phased light curve on the website\footnote{http://keplerebs.villanova.edu/}. KOI-131 is listed as a planet candidate in \cite{batalha12} as well.  The highest out-of-transit peak has a 86 day period for this system. Therefore it is likely that the TTV observed at 114 days period is not related to brightness variations of the host star. We conclude that although this object passes all our test, it is not a strong hot-Jupiter candidate based on the available information.

\noindent {\bf \object{KOI-256}} According to \cite{muirhead12} the host star is a metal-rich M3 dwarf. Their measurement significantly reduced the size of the planet candidate from 25.34 to 5.60\thinspace ${\rm R_E}$. The companion shows a single 41.8 d periodicity in its transit timing. 

\noindent {\bf \object{KOI-823}}  This object fell on the dead $Kepler$ module in Q6. It shows complicated, multipeaked Fourier-spectrum. While some of the peaks are close to peaks seen in the light curve itself (142.9, 59.5, 19.0 d, 1.02 d -  which is the highest out-of-transit peak), others are found to be independent of stellar and sampling effects.

\noindent {\bf \object{KOI-882}} Single-periodic TTV candidate. The long period seen in the TTV (41.9\thinspace d) is unlikely to be disturbed by other relatively long-period periodicities found in the light curve, the closest one is 37.1\thinspace d. 
The highest out-of-transit peak with 3.92\thinspace d period has 7300\thinspace ppm amplitude, and alerts for severe activity effects. However, it is not clear how the long period TTV signal is affected by such a short rotation period.

\subsubsection{Rejected candidates}

\noindent {\bf \object{KOI-412}} Data from Q5 is missing because the target fell on the failed $Kepler$ Module 3. The 
highest peak in the Fourier spectrum of the transit timing data of this object is located exactly at the 
Nyquist-frequency, i.e. twice the orbital period of the planet. We cannot exclude a signal resulting from a resonant 
object with 1:2 mean motion resonance in the system. The amplitude of the TTV is uncertain for the same reason. 

\noindent {\bf \object{KOI-822}} For similar reasons to the case of KOI-412, Q5 data are is missing. Strong 
rotational modulation is seen in the light curve with about half the detected TTV period, as well as its first harmonic, 
so it may influence the transit timing measurements. Therefore we flagged this star as an uncertain case. 

\noindent {\bf \object{KOI-883}} Based on its KIC-parameters, the host star is an early type K dwarf. We 
found two peaks in the Fourier spectrum of the transit times of its companion (6.6\thinspace d and 
9.1\thinspace d). The 6.6\thinspace d peak is uncomfortably close to the alias caused by the 
sampling effect, while the latter coincides with the rotation period derived from its light curve presumably caused by
the presence of a spot or spots. Thus, this object is no longer a candidate showing periodic TTVs. We decided 
to leave it out of Table.~\ref{apptable}.

\noindent {\bf \object{KOI-895}} is a  single-period companion around a solar-mass star with 11.4 d TTV period. The 
stellar rotation, however, creates a periodicity of 5.6 d, which is almost exactly half of the TTV period. This renders 
the main TTV period questionable. 

\noindent {\bf \object{KOI-1003}} This is a monoperiodic TTV candidate with a long but highly significant period (277 
days). The activity level of the star is very high, and it may affect timing, so may result in a TTV. Most important, 
\cite{ofir12} list this object as an eclipsing binary.

\begin{figure}
\centering\includegraphics[height=9.0cm,angle=270]{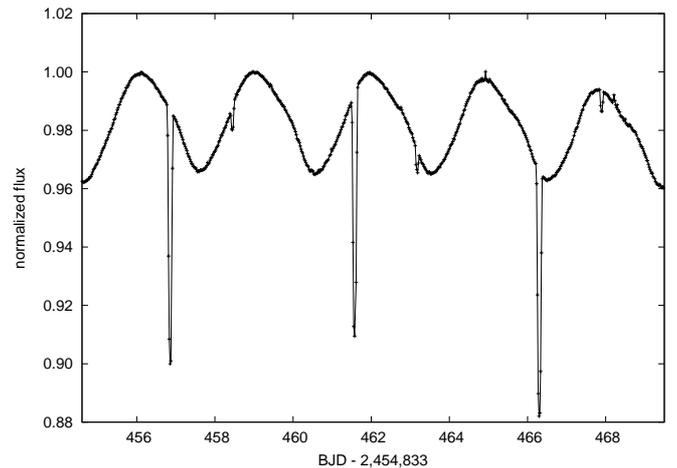}
\caption{A small part of the $Kepler$ light curve of KOI-1152, clearly showing transits and secondary eclipses. The secondary eclipses occur at phase 0.34, indicating an eccentric orbit.}
\label{koi-1152}
\end{figure}

\noindent {\bf \object{KOI-1152}} A larger-than-Jupiter companion orbits an early M1 dwarf \citep{muirhead12}). The star shows a three-day rotation period. Besides the deep transits, the secondary minima (occultation) are 
also easily seen. In addition, the orbit of the companion is obviously eccentric, since the secondary eclipses are detected around 
the 0.34 phase instead of 0.50 (see Fig.~\ref{koi-1152}).  The lagged occultation refers to a minimum eccentricity of 
0.26 (\cite{sterne40}, or more recently Eq. (2) in \cite{dong13}). The circularization time scale for this 
system is $\tau_{circ} \sim$ 120--270 Myr, assuming a planet mass of 1~${\rm M_J}$, a tidal quality factor of 
$10^6$, and other necessary system parameters as derived in KIC and in \cite{batalha12}. The shorter time 
scale assumes an eccentricity of $0.26$, and the longer time scale is the $e\sim 0$ limit. Because the circularization time 
scale is much less than the expected lifetime of an $M_*=0.58 {\rm M_{\sun}}$ star, it is reasonable that the system is older 
than $\tau_{circ}$, and the eccentricity is maintained by a dynamical process. Therefore, this system is a good 
candidate for having an outer companion that perturbs the orbit. \cite{ofir12} classify this object as an eclipsing binary.

\noindent {\bf \object{KOI-1285}} The optical light curve of this multifrequency TTV candidate shows peaks close to the 
53 and 70\thinspace d TTV periodicities. This is an active star, exhibiting an out-of-transit frequency at 1.07
\thinspace d with 1450\thinspace ppm amplitude; however, this peak is far from the detected TTV signal. Based on the  latest release of the list of KOIs, it is a false positive.

\noindent {\bf \object{KOI-1382}} This system shows two very clear periodicites (with 34.4 and 17.4 days), close to, but 
not exactly at, a 1:2 period ratio. The rotation period of the $1.1 M_{\odot}$ host star is 4.8 days. This is another false 
positive. 

\noindent {\bf \object{KOI-1448}} According to \cite{batalha12}, the system consists of a relatively large companion (${\rm R= 24.25\thinspace R_E}$) orbiting a solar-like host star with a relatively strong  11.18\thinspace d period TTV signal. However, lately it has been found to be a false positive. 

\noindent {\bf \object{KOI-1452}} The size of the companion of this star is twice that of the Jupiter. The system exhibits multiple TTV periodicities. The 74.4 and the 58.3 d peaks are close to peaks detected in the Fourier spectrum of the photometric light curve, therefore are ambiguous. The stellar rotation period is estimated to be 1.5 days based on the $Kepler$ light curve. The system survived all our criteria, and it would remain a strong multiperiodic TTV candidate. 
\cite{mazeh13}, however, concludes that it is an eclipsing binary system.

\noindent {\bf \object{KOI-1540}} If indeed a planet, this is one of the largest candidates in our vetted sample with 
almost three Jupiter radius, orbiting a solar-mass star. One of the six significant periods (33 d) is found to be spurious from the light curve analysis. The period with 2.427 \thinspace d is roughly twice as long as the orbital period of the planet (thus the  sampling), which is 1.208 \thinspace d, therefore ambiguous. According to \cite{ofir12} and the latest KOI release it is not a bona fide planetary system. 

\noindent {\bf \object{KOI-1543}} According to our analysis, KOI-1543 is a monoperiodic TTV candidate with a 
3.96\thinspace d orbital and a 97.00 \thinspace d TTV period around a star of $1.06\thinspace {\rm M_{\odot}}$ mass and $0.86\thinspace {\rm R_{\odot}}$ radius. However, it turned out to be a false positive. 

\noindent {\bf \object{KOI-1546}} This object features the richest spectrum of transit timing variations in our sample. Three frequencies (62.6, 357.9, 74.3 d) are uncertain, because these appear in the light curve as well, another ten peaks reach the four  standard deviation level. According to \cite{mazeh13}, it is also an eclipsing binary system.

\section{Interpretation}

Observing periodic TTV variations in hot Jupiters has been completely unexpected in light of the loneliness paradigm, and it 
would imply that one of the following three scenarios acts in systems with hot Jupiters, and affects the transit times:
\begin{itemize}
\item{} systems with transiting hot Jupiters having non-transiting, massive close-in planets on highly inclined orbits; 
\item{} hot Jupiters with significant TTVs host large moons; or
\item{} some other, still unidentified source of periodic TTV is in action.
\end{itemize}

\begin{figure*}
\centering\includegraphics[bb=80 230 430 440,height=4.8cm]{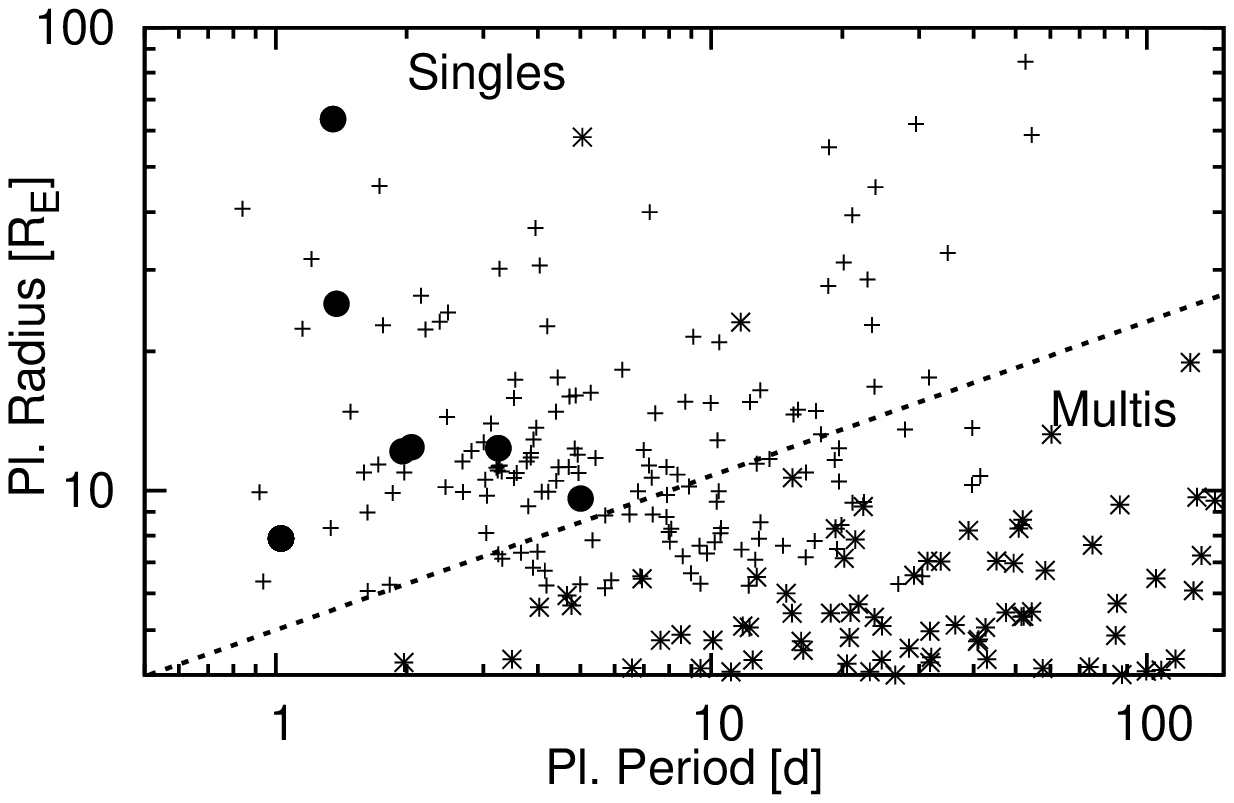}\hskip3mm\includegraphics[bb=80 240 430 450,height=5.2cm]{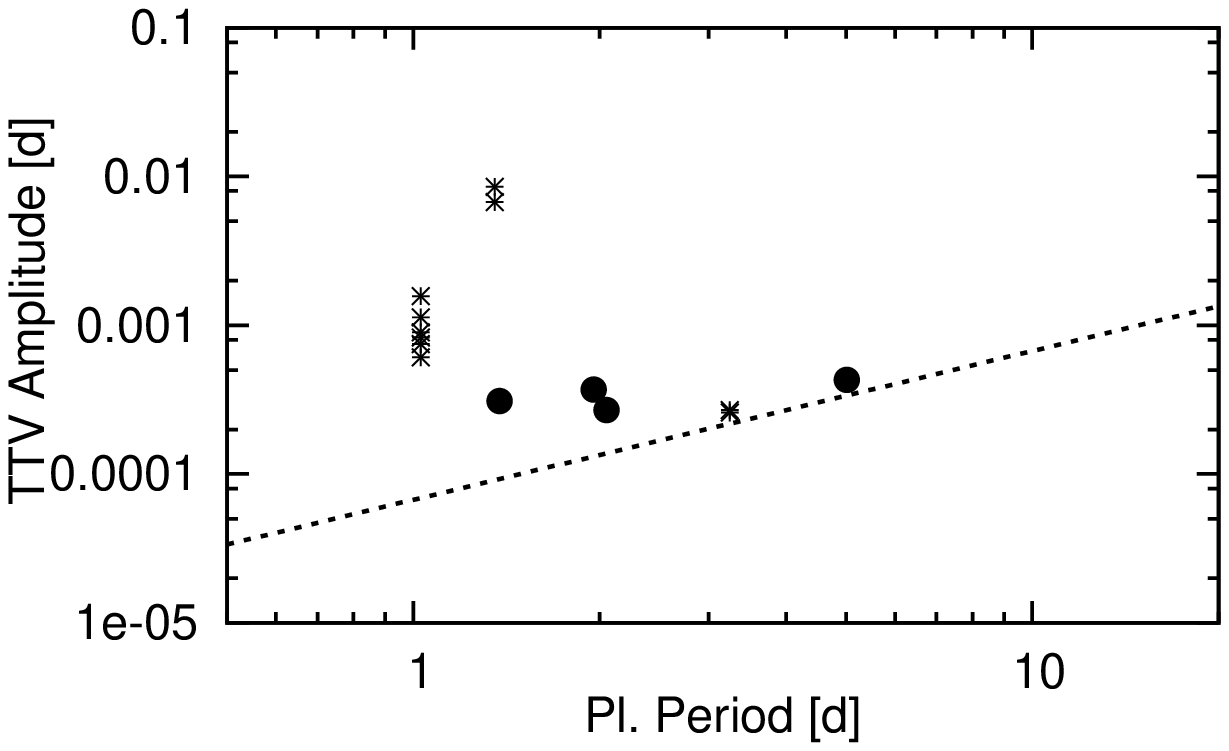}
\caption{{\bf Left:} Test for singleness of TTV hot Jupiters. Stars: multiple systems; crosses: single hot Jupiters; large dots: hot Jupiters with significant TTV. The line shows the upper boundary of the region occupied by known coplanar multi systems. {\bf Right:} Test for moons. Large dots: hot Jupiters with single TTV periodicity; stars: systems with multiple frequencies taken from Table~\ref{table}. The line plots the expected TTV amplitudes assuming 1~$M_J$ planet with a 1~$M_E$ moon orbiting at the Hill radius.}
\label{test}
\end{figure*}
\subsection{Are there more planets out there?}

Hot Jupiters are not necessarily single. If that is the case, periodic TTVs may be tracers of perturbations from additional planets in the system, which are not observed in transits because of the viewing geometry. A known case for a nonsingle hot Jupiter is in \object{HAT-P-13}, hosting a hot Jupiter and a distant massive planet on an eccentric orbit, which is not observed in transits \citep{szabo10}. In the \cite{batalha12} catalog of 2321 planet candidates, there are several examples of hot Jupiters in multiple systems, e.g. KOI-338, a system with two planets, where both are hot Jupiters. KOI-94 consists of four transiting candidates, one hot Jupiter, two hot Neptunes, and a hot super-Earth, KOI-1241 hosts a hot Jupiter and a hot Neptune. \object{WASP-12} is a candidate for a two-planet system, where the TTV signal of the transiting component and a slight modulation of radial velocity data both suggest a second companion \citep{maciejewski11, maciejewski13}.

However, as the lefthand panel of Fig.~\ref{test}. shows, the distribution of hot Jupiter candidates with TTV is very different from the distribution of planets in multiplanet systems. In particular, very few multiple systems exhibit planets above the line $R/R_E > 5 \sqrt[3]{P/day}$. On the other hand, all TTV hot Jupiters reside above this line, therefore they are compatible with the distribution of non-TTV, single hot Jupiters. One could argue that massive, nontransiting companions on an inclined orbit can perturb the transiting planet, but in this case, the planets must also be close-in types to support the median detected TTV period of 50 days, and these planets would also show up in radial velocity planet-searching surveys.  This point is discussed 
in detail in \cite{steffen12c}. We note that these systems would look very different from, e.g., \object{HAT-P-13}.

\subsection{Are there exomoons out there?}

Another -- however, still unconfirmed -- explanation of TTVs is the presence of moons pumping TTVs \citep{sartoretti99,szabo06,simon07,kipping09}. The expected rate of a TTV can be estimated if we assume that the satellite has no sign in the light curves, but that it only affects the transit time of the planet. In this case, the TTV's full amplitude is $a_s M_s P(\pi a M_p)^{-1}$, where $a_s$ and $a$ are the semi-major axis of the satellite and the planet, ${\rm M_s}$ and ${\rm M_p}$ are their mass, respectively, and $P$ is the orbital period (Sartoretti and Schneider, 1999). For an order-of-magnitude estimate of the expected effects, we assigned 1~${\rm M_J}$ mass to the planet, 1~${\rm M_E}$ mass to the exomoon, and assumed $a_s=a_H$, i.e., the moon orbits at the Hill-radius. With these assumptions, TTV caused by a moon can be calculated as a function of the orbital period of the planet.

In the righthand panel of Fig.~\ref{test}, we plot the measured TTV amplitudes for the hot Jupiters. Systems with multiple 
TTV periods are plotted with different symbols, and they follow the same trend as the monoperiodic systems. The linear 
model overplots the expected TTV amplitude of a $Jupiter+Earth$ system. We conclude that the distribution of TTV points above an orbital period of two days follows a similar trend as the model, but for shorter orbital periods, TTVs significantly exceed  this prediction. However, TTV can be underestimated by a factor of 2--5 based on the \cite{sartoretti99}  
assumptions, especially when the moons are quite large (see \cite{szabo06}, \cite{simon07}), 
and the detected TTVs may be explained with less massive exomoons.

An argument against the exomoon scenario is that at least some fraction of hot Jupiters should have distant and massive moons to explain our findings. Although the existence of large (M$>$1~$M_E$) exomoons is an attractive possibility, 
an appropriate -- still unknown -- scenario needs to address the formation of large exomoons 
around hot Jupiters. Besides the origin and prevalence of such systems, any interpretation also faces the question of stability. Hot Jupiters with a moon suffer tidal forces from both the star and from the close-in, massive exomoon. As a consequence, the rotation rate of the planet evolves, and the orbital distance of the exomoon can swing very significantly, either toward spiral-in, or escape, or oscillation, or chaotic behavior \citep{barnes02}. Considering this argument, one could raise a reasonable doubt about the assumption of massive exomoons around a significant fraction of hot Jupiters, though it may be possible to observe such moon-planet systems that are young or formed in 
exotic scenarios, such as capture.

\subsection{Comparison with independent studies}

 During the submission process of this paper, \cite{kipping13} published a set of $Kepler$ planet candidates, seeking for scenarios that are compatible with moons. The two studies are based on different methods: Kipping et al. analyzed TTV distributions, and here we applied a period analysis. The two candidate lists are also disjointed, because \cite{kipping13} restricted their analysis to candidates smaller than six Earth-radii. There are nevertheless prominent similarities between the two studies: neither we nor \cite{kipping13} succeeded in firmly concluding that TTV signals have a clear dynamical origin, although a few strong candidates survived all tests.

\cite{mazeh13} have followed a similar approach to ours, and during the revision of this paper published another list of candidates based on an extended time base up to Q12. Some of our candidates (especially the multifrequency TTV ones) are confirmed, a few are found to be false positives, but others are not mentioned as significant detections by \cite{mazeh13}. 

\cite{steffen12c}  conducted an independent analysis to find TTV variations and a photometric search for additional transits in $Kepler$ systems containing hot Jupiters based on the same data we used (Q0--Q6). They conclude that there is a significant difference in the presence of additional components between hot Jupiters and other populations (warm Jupiters, hot Neptune candidates), which may be the consequence of their different dynamical histories. They failed to identify 
those candidates that were retained as showing short-term or other significant (most probably multifrequency or nonperiodic) TTV variations by \cite{mazeh13}. (One such example is KOI-882 in our final list of seven TTV candidates.) The reason \cite{steffen12c} focused on long-term TTV periodicities is that is what one physically expects for near-resonant configurations. 
The sensitivity to nonresonant perturbations is dramatically reduced. Thus, it is a priori unlikely to find dynamically 
induced TTV signals with short periods.

The diverging results of these studies  are alerting signals that even currently available data are insufficient for firm detections (especially for long-period variations). In spite of this, testing methods and seeking TTV candidates are in the forefront of hot Jupiter studies.

\subsection{Summary and conclusions}

 In this paper, we have shown that
\begin{itemize}
\item{} Equidistant sampling leads to periodic, artificial TTV signals in $Kepler$ exoplanet photometry, and data analysis must be cleared for the stroboscopic effect;
\item{} Stellar rotation and activity can also be linked to TTV signals and, in many cases, can mimic a virtual transit timing variation that is still nonphysical;
\item{} In the case of three Kepler targets (KOI-186, 897, 977), a periodic TTV signal is measured that is a nonstroboscopic signal from relatively anactive stars, and might have a dynamical origin.
\end{itemize}

\cite{santerne12} have recently estimated that the rate of false positives among short-period Jupiters
in the $Kepler$ sample can be as high as 34.8\% $\pm$ 6.5\%. In the framework of their 
radial velocity follow-up survey, none of our candidates were confirmed as planets, 
but a few were rejected as discussed in Sec.~\ref{individual}. Previously, 
\cite{morton11} estimate a rate that is closer to 5-10 \%. Depending on which value is taken, 
the number of nonplanetary candidates may be between one and six in our sample of nineteen  hot Jupiter TTV candidates, 
even if brown dwarfs are permitted. More follow-up observations are clearly needed to settle this problem. 
It is worth noting that recently nine in this sample were found to be eclipsing binaries or other false positives 
\citep{ofir12, mazeh13}.

 It is remarkable that except for KOI-977, which should be an evolved star orbited by the largest substellar companion in our sample as we discussed in Sec.~\ref{individual}, and for KOI-256 which is an M dwarf hosting a large Jupiter-type planet, the remaining five of our best candidates in Table~\ref{table} are very similar, both in the 
parameters of their host stars and in the parameters of their transiting hot Jupiters. The stellar masses are slightly above solar (ranging form 0.93 to 1.13 $\rm M_{\odot}$), while the corresponding radii (0.97-1.21 $\rm R_{\odot}$) suggest main-sequence objects. 
The size of the companions point to `canonical Jupiters' in the 7.9-12.4 Earth-radius range. The orbital periods are between 1.0 and 5.0 days. If the seemingly periodic TTV indeed has a dynamical origin, then maybe it is not a coincidence that we find objects with such similar architecture and so any explanation should also reflect this tendency.  We note that the TTV amplitudes of most of the best candidates are considerably smaller than the TTV amplitudes of most systems confirmed or characterized by TTVs (eg. \cite{holman10,steffen13}).

In many cases we suspect that the Q0-Q6 data set is simply not long enough to assess the real TTV behavior 
of our candidates.
Longer observations and ground-based follow-ups can confirm the nature of the TTV-positive hot Jupiters. With the help of further $Kepler$ short cadence observations, or even with accurate transit photometer with large telescopes applying much denser sampling than $Kepler$, sampling effects in long-cadence data can be eliminated. If the physical nature of the observed hot Jupiter TTVs can be confidently confirmed, only dynamical explanations survive: additional planets on exotic orbits or the exomoons.

\section*{Acknowledgments}
We thank the referee for providing constructive comments and helping improve the contents of this paper.
This project has been supported by the Hungarian OTKA Grants K76816, K83790, K104607, the HUMAN MB08C 81013 
grant of the MAG Zrt., the ``Lend\"ulet-2009 Young Researchers'' Program of the Hungarian Academy of Sciences 
and by the City of Szombathely under agreement No. S-11-1027. RSz and GyMSz were supported by the J\'anos Bolyai Research Scholarship of the Hungarian Academy of Sciences. Eric Ford and Jason Steffen are acknowledged for their help with the electronic table of $Kepler$ TTV measurements. We thank Daniel Fabrycky for insightful comments on an early version of the manuscript. We are grateful to the Kepler Team for making the light curves public and thank all those who worked hard 
to make this mission a success. This research made use of the NASA Exoplanet Archive, which is operated by the California Institute of Technology, under contract with the National Aeronautics and Space Administration under the Exoplanet Exploration Program.

\bibliographystyle{aa} 
\bibliography{singles_SzR.bib}


\begin{appendix}
\section{On-line table}

\longtab{1}{
\begin{longtable}{rlllllrrrc}

\caption{Rejected periodic TTV detections in our $Kepler$ hot-Jupiter candidate sample. The vetting was based either on  nongravitational processes at work (e.g. stellar rotation) that affect the transit time determination or on the 
reported improved status (false positive and/or eclipsing binary) of the candidates. Periods in italics are ambiguous, based on the analysis of the stellar optical light curves of the candidates. Stroboscopic periods are omitted. In the last column we refer to the cause  of rejection FP: a false positive based on the 2013 January updated release of Kepler Objects of Interest.}\\
\label{apptable}
KOI &  $R/R_E$    & P (d)     &$M_*$& $T_{eff}$ & $R_*$ & TTV per. (d) &  Amp. (d) & Sign. lev. (s.t.d.) & Remarks \\
\hline
\endfirsthead
\caption{continued.}\\
KOI &  $R/R_E$    & P (d)     &$M_*$& $T_{eff}$ & $R_*$ & TTV per. (d) &  Amp. (min) & Sign. lev. (s.t.d.) & Remarks \\
\hline
\endhead

 412.01  & 6.72   & 4.1470197  & 1.09 & 5584&  1.17 & {\it   8.29}      &{\it $>$1.4544}&{\it $>$4.1}  & rot, 1\footnotetext[1]{the peak close to the Nyquist-freq., i.e. $2\thinspace P_{orb}$ } \\
 822.01  &9.79    & 7.9193704  & 1    & 5458&  0.74 & {\it  24.523629}  &{\it 1.7424} &{\it  7.4}  &  rot\\
 895.01  & 10.51  & 4.4094114  &1.04 & 5436  &0.93  & {\it 11.436544} &{\it 0.432} &{\it  5.1} & rot \\
1003.01  & 10.84  & 8.3605703  &0.96 & 5126 & 0.8   &  277.700639 &2.9808 & 9.6 & FP, EB, 2\footnotetext[2]{Classified as eclipsing binary by \cite{ofir12}}\\
1152.01  & 15.97  & 4.7222521  &0.58 & 4069 & 0.55  &{\it 20.868549} &{\it 0.7056} &{\it  9.9} &  FP, EB, 2 \\
                  &&&&&                             &  11.810977  &0.5616 & 7.5 &   \\
1285.01  &  6.36  & 0.9374439  &0.98 & 5278 & 0.83  & 374.812594  &5.8608 &65.1 &  FP \\   
                  &&&&&                             & 106.826193  &1.9440 &20.3 &  \\
                  &&&&&                             & 162.786912  &1.4976 &15.2 &  \\
                  &&&&&                             &{\it 53.455926} &{\it 0.5616} &{\it  4.5}  &  \\
                  &&&&&                             &{\it 70.229651} &{\it 0.5472} &{\it  4.3}  &  \\
1382.01  & 22.64  & 4.2023359  &1.1  & 5921 & 1.07  &  34.449497  &1.4688 &17.8  & FP \\
                  &&&&&                             &  17.397961  &0.9792 &11.3  &  \\
1452.01  & 22.39  & 1.1522169  &1.27 & 6834 & 1.76  &  30.290180  &2.1312 &37.6  & EB, 3\footnotetext[3]{Classified as eclipsing binary by \cite{mazeh13}}\\
                  &&&&&                             &   6.308352  &1.8144 &31.7  &  \\
                  &&&&&                             &   4.802105  &1.2960 &22.1  &  \\
                  &&&&&                             &{\it 74.432453}  &{\it 1.2672} &{\it 21.6}  &  \\
                  &&&&&                             & 144.927536  &1.1952 &20.3  &  \\
                  &&&&&                             &  44.359668  &0.5328 & 8.1  &  \\
                  &&&&&                             &{\it 58.309038}  &{\it 0.5184} &{\it  7.8}  &  \\
                  &&&&&                             & 192.864031  &0.4608 & 6.7  &  \\
                  &&&&&                             &  50.584248  &0.4320 & 6.2  & \\
1448.01  & 24.25  & 2.4865874  &1.08 & 5658 & 1.17  &  11.177805  &0.9360 & 8.4  & FP \\
1540.01  & 31.64  & 1.2078535  &1    & 5390 & 0.77  & 390.015601  &0.6192 & 9.5  & FP, EB, 2 \\
         &&&&&                                      &  55.962841  &0.5904 & 8.9  & \\
         &&&&&                                      &{\it 2.426855}  &{\it 0.5904} &{\it  8.9}  & \\
         &&&&&                                      &  44.640864  &0.5472 & 8.1  & \\
         &&&&&                                      &   2.485046  &0.4176 & 5.8  & \\
         &&&&&                                      &{\it 32.99894}  &{\it 0.3888} &{\it  5.3}  & \\
1543.01  & 13.69  & 3.9643337  &1.06 & 5821 & 0.87  &   96.99321  &0.4464 & 9.1  &  FP\\ 
1546.01  &  9.92  & 0.9175471  &0.93 & 5505 & 0.86  &  141.74344  &5.3136 &77.6  & EB, 3\\
         &&&&&                                      &{\it 62.63702}  &{\it 2.7648} &{\it 39.5}  & \\
         &&&&&                                      &{\it 357.90981} &{\it 2.2752} &{\it 32.2}  & \\
         &&&&&                                      &{\it 74.32181}  &{\it 1.4976} &{\it 20.6}  & \\
         &&&&&                                      &  230.36167  &1.4544 &20.0  & \\
         &&&&&                                      &  107.71219  &1.3968 &19.1  & \\
         &&&&&                                      &   44.161809 &0.9216 &12.0  & \\
         &&&&&                                      &   55.694793 &0.6912 & 8.6  & \\
         &&&&&                                      &   33.392326 &0.6480 & 7.9  & \\
         &&&&&                                      &   86.625087 &0.5472 & 6.4  & \\
         &&&&&                                      &   36.549708 &0.4896 & 5.5  & \\
         &&&&&                                      &   10.897400 &0.4464 & 4.9  & \\
         &&&&&                                      &   21.113973 &0.4176 & 4.5  & \\   

\end{longtable}
}

\end{appendix}

\end{document}